\documentclass[preprint]{aastex}
\pdfoutput = 1          
\usepackage{natbib}     
\usepackage{color}
\newcommand{\thpar}{$\theta_{\rm par}$}
\newcommand{\thcr}{$\theta_{\rm CR}$}
\newcommand{\ipa}{$\theta_{\rm IPA}$}
\newcommand{\iaa}{$\theta_{\rm IAA}$}

\newcommand\note[1]{\textbf{\color{red}#1}}   
\newcommand\todo[1]{\textbf{\noindent \color{red} $\square$ #1}}
\newcommand\new[1]{{\color{blue}#1}}          
\renewcommand\note[1]{\hspace{-1sp}}
\renewcommand\todo[1]{\hspace{-1sp}}
\renewcommand\new[1]{\color{black}#1}

\shorttitle{NICI Astrometric Calibration}
\shortauthors{Hayward et al.}

\begin{document}

\title{Astrometric Calibration of the Gemini NICI Planet-Finding Campaign\footnotemark[1]}

\author{Thomas L. Hayward\altaffilmark{2}, Beth A.
  Biller\altaffilmark{3,4}, Michael C. Liu\altaffilmark{5}, Eric L.
  Nielsen\altaffilmark{5}, Zahed Wahhaj\altaffilmark{6}, Mark
  Chun\altaffilmark{5}, Christ Ftaclas\altaffilmark{5}, Markus
  Hartung\altaffilmark{2}, and Douglas W. Toomey\altaffilmark{7}}

\footnotetext[1]{Based on observations obtained at the Gemini
  Observatory, which is operated by the Association of Universities
  for Research in Astronomy, Inc. under a cooperative agreement with
  the NSF on behalf of the Gemini partnership: the National Science
  Foundation (United States), the National Research Council (Canada),
  CONICYT (Chile), the Australian Research Council (Australia),
  Minist\'{e}rio da Ci\^{e}ncia, Tecnologia e Inova\c{c}\~{a}o (Brazil), and
  Ministerio de Ciencia, Tecnolog\'{i}a e Innovaci\'{o}n Productiva
  (Argentina).}

\altaffiltext{2}{Gemini South Observatory, AURA/Casilla 603, La Serena, Chile}
\altaffiltext{3}{Institute for Astronomy, University of Edinburgh, Blackford Hill, Edinburgh, EH9 3HJ, UK}
\altaffiltext{4}{Formerly at Max-Planck-Institut f\"ur Astronomie, K\"onigstuhl 17, 69117 Heidelberg, Germany}
\altaffiltext{5}{Institute for Astronomy, University of Hawaii, 2680 Woodlawn Drive, Honolulu, HI 96822, USA}
\altaffiltext{6}{European Southern Observatory, Alonso de Cordova 3107, Vitacura, Casilla 19001, Santiago, Chile}
\altaffiltext{7}{Mauna Kea Infrared LLC, 21 Pookela St, Hilo, HI 96720}

\begin{abstract}
  We describe the astrometric calibration of the Gemini NICI
  Planet-Finding Campaign.  The Campaign requires a relative
  astrometric accuracy of \new{$\approx 20$~mas} across multi-year
  timescales in order to distinguish true companions from background
  stars by verifying common proper motion and parallax with their
  parent stars.  The calibration consists of a correction for
  instrumental optical image distortion, plus on-sky imaging of
  astrometric fields to determine the pixel scale and image
  orientation.  We achieve an accuracy of \new{$\lesssim 7$~mas
    between the center and edge of the 18$''$ NICI field}, meeting the
  20~mas requirement.  Most of the Campaign data in the Gemini Science
  Archive are accurate to this level but we identify a number of
  anomalies \new{and} present methods to correct the errors.
\end{abstract}


\section{Introduction}
\label{sec-intro}
\nobreak
The Gemini NICI Planet-Finding Campaign was a direct-imaging survey of
\new{about} 250 nearby stars for substellar and planetary-mass companions
conducted at Gemini South Observatory between 2008 and 2012
\citep{2010SPIE.7736E..53L}.  The Campaign used the Near-Infrared
Coronagraphic Imager \citep[NICI,][]{2008SPIE.7015E..49C}, which
combines adaptive optics, coronagraphy, angular differential imaging
\citep{2004Sci...305.1442L, 2006ApJ...641..556M}, and dual-channel
methane-band infrared imaging to achieve an H-band contrast
detection limit of 14.4~magnitudes at 1$''$ radius
\citep{2013ApJ...779...80W}.  The principal scientific results have
been published by 
\citet{2010ApJ...720L..82B}, 
\citet{2011ApJ...729..139W},
\citet{2012ApJ...750...53N},
\citet{2013ApJ...773..179W},
\citet{2013ApJ...776....4N}, and 
\citet{2013ApJ...777..160B}; 
the pipeline processing algorithm is described by
\citet{2013ApJ...779...80W}.

The multi-epoch imaging data acquired for the Campaign require
accurate astrometric calibration in order to detect common proper
motion and parallax between the host star and any candidate companions and
therefore distinguish true companions from background stars.  In this
paper, we describe the process used to establish the alignment of the NICI science
detectors relative to the celestial coordinate system and calibrate the World
Coordinate System (WCS) contained in the data headers.

\section{Instrument Design and Data Format}
\label{sec-instdesign}
\nobreak
NICI consists of two parts: an adaptive optics (AO) system and a
dual-channel science camera \citep{2008SPIE.7015E..49C}.  The
AO-corrected wavefront entering the science camera first passes
through a focal plane coronagraph mask and a pupil/spider mask, then is divided
between the two science channels by the
{\it Dichroic Wheel} (DW).  Light transmitted through the DW enters the
{\it Red} channel while reflected light enters the {\it Blue}. 
The DW can be set to one of the following elements: {\it H-50/50}
beamsplitter, which directs 50\% of H-band (1.65~\micron) light to
each science channel; {\it H/K Dichroic}, which divides light between the H and
K (2.2~\micron) bands; {\it Mirror}, which reflects all light to the
Blue channel; or {\it Open}, which passes all light to the Red channel.

After the DW, each science channel contains a filter wheel, reimaging
optics, and an ALADDIN~II 1024$\times$1024 InSb array detector.  The
Campaign most frequently used 4\% bandpass filters centered in and out
of the $\lambda = 1.63~\mu$m methane absorption feature (named
CH4-H4\%L and CH4-H4\%S, respectively) to search for methane-bearing
planets very close to the primary star, and a broadband H filter for
deeper searches at larger separations.  The on-sky pixel size is
$\approx 18$~milli-arcseconds (mas) and the field size is $\approx
18\farcs4 \times 18\farcs4$.  The mapping of the sky onto each of the
two science detectors is different and changes with the DW element and
filter.

NICI data are processed by the Gemini Data-Handling System (DHS) and
are written to Multi-Extension FITS (MEF) files
\citep{1981A&AS...44..363W, 2010A&A...524A..42P} which contain a
Primary Header followed by two extensions, one for each of the two
science channels.  The header for each extension contains its own WCS
specifying the astrometric parameters for that channel
\citep{2002A&A...395.1061G}.  The calibration of the WCS for each channel is a
principal subject of this work.

Details on the data format, WCS, and other header data are specified
in the Appendix, and more information on NICI is available at
\url{http://www.gemini.edu/sciops/instruments/nici}.

\section{Astrometric Requirements}
\label{sec-astreq}
\nobreak
The astrometric calibration requirements for the NICI Planet-Finding
Campaign arise from three aspects of the observing strategy.

First, most Campaign data were taken in a mode called 
{\it Angular Differential Imaging} 
\citep[ADI,][]{2004Sci...305.1442L, 2006ApJ...641..556M}, 
in which the Cassegrain rotator is held fixed and the sky 
allowed to rotate on the science detectors so that astronomical
objects surrounding the primary star under observation move relative
to the fixed speckles in the NICI point-spread function (PSF).  
The individual $t \approx 1$ minute exposures taken over a range
of field angles (in some cases $> 90^\circ$) were 
derotated and registered before being co-added by the 
pipeline-processing software.  To achieve accurate alignment over the 
entire field, image distortion must be corrected before derotation.

Second, targets were imaged simultaneously in the two science channels
in and out of the $\lambda = 1.63~\mu$m methane absorption feature, in a mode
known as {\it Angular Spectral Differential Imaging} (ASDI), to
suppress the stellar continuum further and enhance potential
methane-bearing planets.  The distortion, rotation and scale
difference between the two channels must be corrected before the
pipeline can accurately register and subtract the two images,
especially when they are added in order to find non-methane-bearing
planets.  In practice, sufficient accuracy can be achieved because
these differences are static properties of the instrument that can be
measured once and applied to all data.

Third, the Campaign strategy required measuring the relative proper
motion (PM) between a target star and any candidate companions
detected in the surrounding field in order to distinguish true
companions (with common PM) from background stars.  Because only the
relative PM is required, we are not concerned with absolute astrometry
(i.e. the absolute celestial coordinate zero points), but only with
the image distortion, field rotation, and pixel scale.  \new{In
order to assess proper motion and parallax robustly, this calibration
had to be maintained over multi-year timescales}, during which time NICI
was dismounted and remounted on the telescope, sometimes on different
instrument ports with different numbers of reflections in the optical
path to the instrument.  Therefore, the astrometric calibration
required observations of astrometric fields during each instrument
mounting.  In addition, because a field observed at multiple epochs was
not typically observed at the same hour angle each time, the field
rotation was often significantly different, again requiring correction of the
image distortion before the two datasets could be compared.

The required astrometric accuracy is dictated by the PM of the target
stars.  In the Campaign's target list of nearby stars, the lowest PMs
are \new{ $\approx 20$~mas yr$^{-1}$, or $\approx 1$~NICI pixel yr$^{-1}$};
most are at least a few times larger.  \new{An accuracy of 1~pixel
  from the center of the NICI field (the usual location of a primary
  star) to the edge} therefore permits most candidate companions to be
checked for common PM within one year, especially since accuracy near
the field center will be considerably better.

For the pixel scale, 1~pixel out of \new{512 is 0.2\%}, while a
rotation error of \new{$0\fdg112$ corresponds to 1~pixel across half the
field.}  These constraints will serve as guidelines throughout this
paper.

\section{Astrometric Calibration}
\label{sec-astromcalib}
\nobreak
Our astrometric calibration process has three principal components:

\begin{enumerate}
\item correcting the image distortion introduced by the instrument optics;
\item measuring the field rotation and setting the instrument alignment angle;
\item constructing the WCS.
\end{enumerate}

\subsection{Image Distortion}
\label{sec-distortion}
\nobreak
The two NICI channels suffer from image distortion arising from the
off-axis reflective optics in both the AO system and science camera.
To measure the distortion, we imaged a pinhole grid mask mounted in
the {\em Fiber Optic Calibration Source} (FOCS), which can be deployed
into the telescope focal plane at the entrance to the AO
system.  The FOCS mask distortion at the science detectors represents
the combined distortions of the AO system and science camera optics.
\new{We preferred to use the mask rather than an astronomical field due
to the grid spots' very high signal-to-noise, their uniform size and
distribution over almost the entire field, and the freedom from seeing
and anisoplanatism effects.}

\begin{figure}[p]
\figurenum{1}
\plottwo{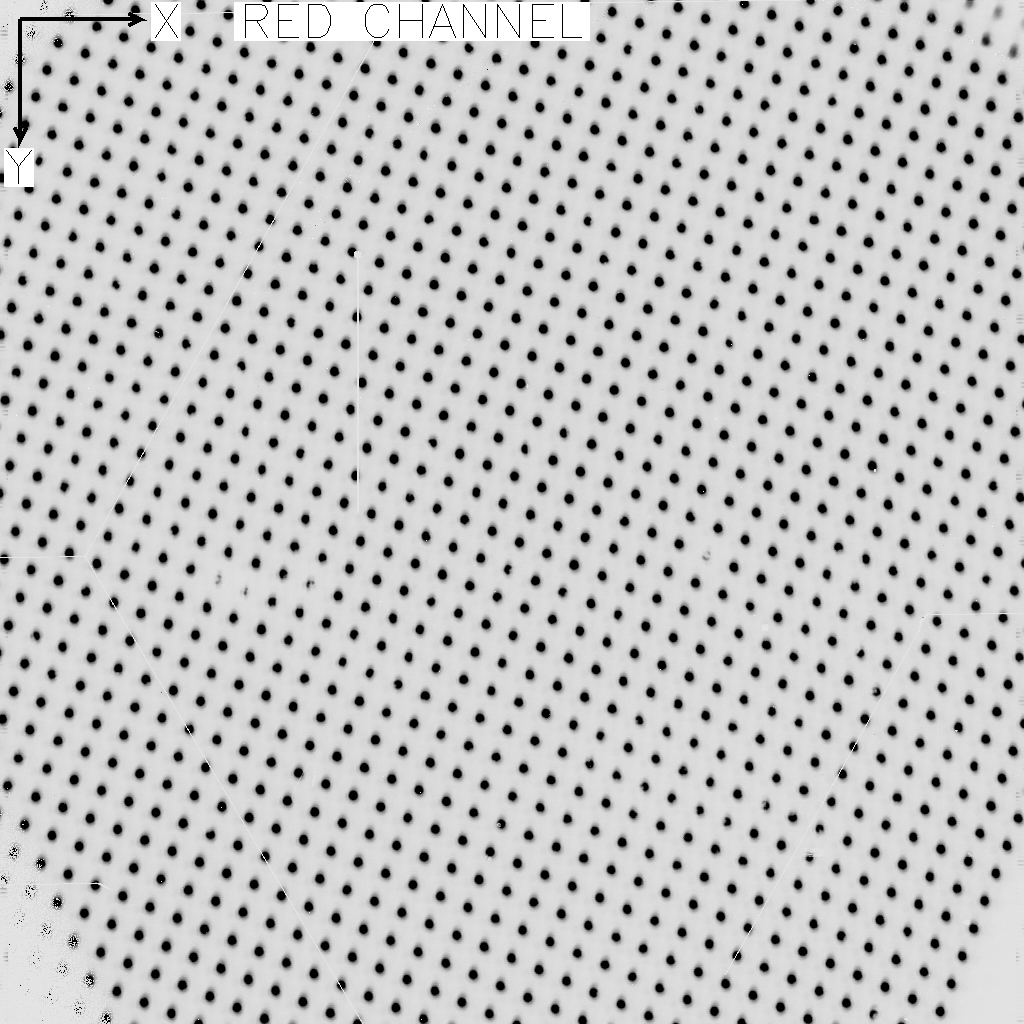}{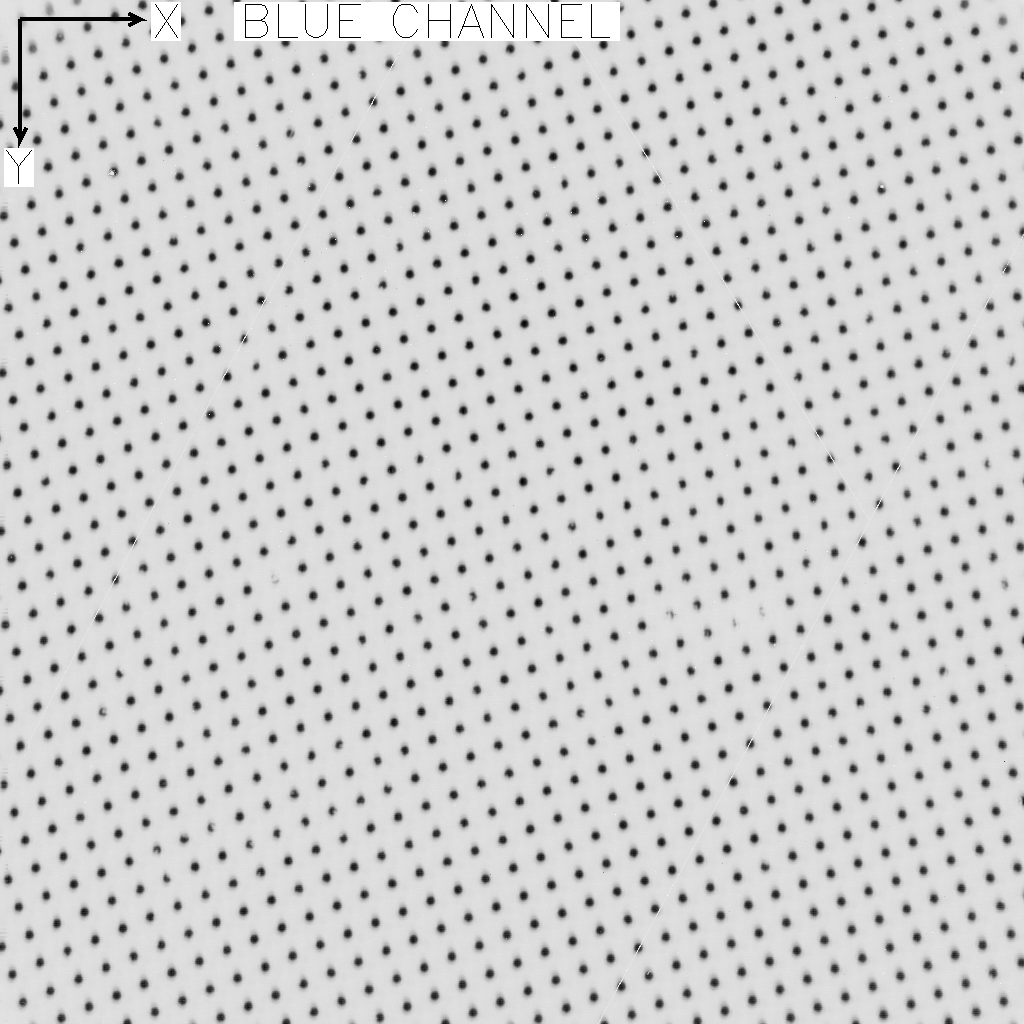}
\caption{NICI FOCS Grid Mask images for the Red (left) and Blue
  (right) channels.  All images in this article are shown with NICI's
  default FITS (row, column) ordering convention in which pixel (1,1) is 
  in the upper left corner and the (X,Y) directions are as indicated in
  the figure.  In this left-handed coordinate system, a positive 
  rotation is clockwise.\label{fig-gridmask}
}
\end{figure}

FOCS Grid Mask images are shown in Figure~\ref{fig-gridmask}.  The
grid is rotated by $22\fdg3$ relative to the detector.  The
Blue channel image is reflected left-right relative to the Red due to
the extra reflection at the DW optic in the path to the Blue channel.

\subsubsection{Fitting Procedure}
\label{sec-fitproc}
\nobreak
To determine the distortion, a grid mask image is first
background-subtracted and the centroid of each spot
measured.  A synthetic rectilinear grid is then fitted to the spot
positions within 256 pixels of the image center, the region
of lowest distortion, by minimizing the $\chi^2$ statistic
\begin{equation}
\label{eq-chisq}
\chi^2 = \sum_{i=1}^{N} \frac{|{\bf r}_i - {\bf r}[x_i(x_0,y_0,d,\theta),\, y_i(x_0,y_0,d,\theta)]|^2}{\sigma_i^2},
\end{equation}
where ${\bf r}_i$ is the measured centroid of spot $i$,
${\bf r}[x_i(x_0,y_0,d,\theta),\, y_i(x_0,y_0,d,\theta)]$ are
the spot positions in the synthetic grid, and 
$\sigma_i$ is the uncertainty of the centroid measurement.  
There are four free parameters: $x_0$ and $y_0$ represent the overall 
grid position, $d$ is the spot separation in pixels (assumed to be 
the same in X and Y), and $\theta$ is the grid rotation.  
The $\chi^2$ value is minimized using the Nelder-Mead simplex method
\citep{1965TCJ..7..308P,Press:1992:NRC:148286} as implemented in the
IDL built-in routine {\tt AMOEBA} (Exelis Visual Information Solutions, 
Boulder, Colorado).  \new{The simplex algorithm begins by computing $\chi^2$ for a set of
trial parameters, then finds the trajectory through parameter space
that steadily reduces $\chi^2$ until a minimum is reached.}  The uncertainty $\sigma$ is 
set to 1 for all points because we are only interested in the best-fit
rectangular grid at this first stage, not the positional uncertainties.

\new{We used the IDL routine {\tt POLY\_2D} to transform the images
  geometrically to correct the distortion.  The transformation is
  defined by polynomials of degree $N$:
\begin{equation}
\label{eq-distrans-x}
x' = \sum_{i=0}^{N}\sum_{j=0}^{N} \, P_{i,j} \, x^j \, y^i
\end{equation}
\begin{equation}
\label{eq-distrans-y}
y' = \sum_{i=0}^{N}\sum_{j=0}^{N} \, Q_{i,j} \, x^j \, y^i
\end{equation}
where $x$ and $y$ are the initial pixel coordinates and $x'$ and $y'$
are the transformed coordinates.  $P_{i,j}$ and $Q_{i,j}$ are
coefficients determined from a least-squares fit performed by the
companion routine {\tt POLY\_WARP}, based on the measured grid mask
($x, y$) positions and the rotated rectilinear grid ($x', y'$)
positions.  After computing ($x', y'$) for all the pixels, 
{\tt POLY\_2D} generates a transformed image using cubic convolution.}
The residual errors of the distortion correction were computed by applying
the correction to the grid mask images, then measuring the difference
between the rectilinear points and the mask spots.

\subsubsection{Distortion Corrections}
\label{sec-distcorr}
\nobreak
We derived distortion corrections for six pairs of images taken in
five different instrument configurations as shown in
Table~\ref{tab-dcparams}.  By experimentation, we determined that a
fifth-order ($N = 5$) fit had a residual error of 0.5 -- 0.6 pixels
root-mean-square (rms) for both channels, and that higher-order
polynomials did not reduce the error.  The initial distortion is as
high as 12~pixels in the array corners.

Given the high signal-to-noise images of the spots, the uncertainty of
the individual centroid measurements is much smaller than 0.5 pixels,
so the quality of the fits is limited by some source of systematic
error.  (In statistical terms, if $\sigma$ is set to 0.1 pixel, the
resultant $\chi^2$ per degree of freedom of the perfect grid relative
to the distortion-corrected images is $\approx 100$, much higher than
the value of 1 expected when random errors dominate.)  The main source
of systematic error is most likely the use of a polynomial to
approximate the distortion; an optical raytrace model may be required
to improve the fit.  The accuracy of the polynomial fit is within our
1~pixel goal, so we accept it as satisfactory for the Campaign.

We fitted six different images to evaluate the consistency of the
distortion across multiple instrument configurations.  The consistency
is a concern because the grid position on the Blue channel array
varies by up to a few dozen pixels (Table~\ref{tab-spotpos}) due to
the slightly different mounting angles of the optical elements in the
Dichroic Wheel, and in some cases the failure of DW to seat
consistently in its detent.  The Blue channel variations are much
larger than the Red because the reflected beam from DW is deviated
more by optical element tilts than the transmitted beam.  For the
H-50/50 beamsplitter, the Blue channel ($X, Y$) positions vary by up to
25 pixels between different positionings (two extremes ``min''
and ``max'' are listed in Table~\ref{tab-spotpos}); evidently the DW
does not fully settle into its detent at the H-50/50 position.  Such
large variations are not detected for the other DW elements.

Across the six datasets, there are only small variations in each of
the parameters for scale and rotation (see Table~\ref{tab-dcparams}).
The spot separations vary by 0.05\% for the Red channel and 0.10\% for
the Blue, while the rotations are consistent to $0\fdg096$ for the Red
channel and $0\fdg088$ for the Blue.  If the Open position is omitted,
a reasonable action given that DW substrates probably introduce
aberrations into the transmitted beam that would
be absent for the Open position, the Red channel rotation range is
only $0\fdg027$.

The Blue channel exhibits DW-dependent rotations which appear to be
correlated with the largest position offsets listed in
Table~\ref{tab-spotpos}, indicating that the irreproducible positions
of the DW are causing both effects.  The rotations are smaller than
our $0\fdg1$ precision goal, and not much larger than the rms error of
the fit, but nevertheless appear to be systematic.

The fitting results do reveal significant differences in scale and
rotation between the Red and Blue channels.  How these differences are
managed is described in the WCS section below.

To evaluate distortion variations between the instrument
configurations, we applied the correction for the H-50/50 + CH4-H4\%
images to the other four datasets.  Table~\ref{tab-dch5050} shows that
in all cases the errors are $< 1$~pixel rms, or $\leq 0.1\%$.

\begin{figure}[p]
\figurenum{2}
\plotone{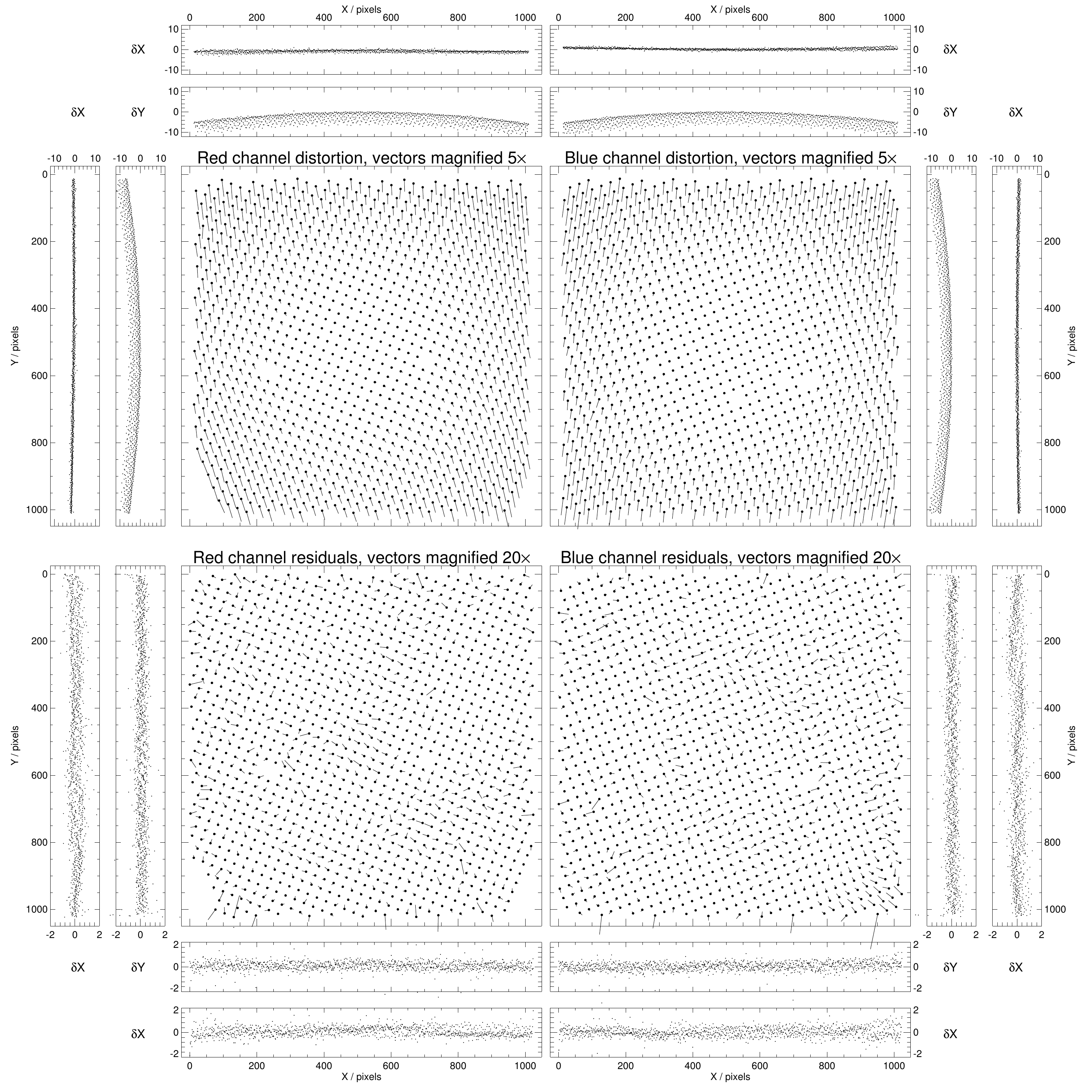}
\caption{\new{Distortion maps for Red and Blue channels with the
    H-50/50 and CH4-H4\% filters, plotted in a format similar to
    \citet{2014A&A...563A..80L}.  The main plots in the top row
    display the original distortion vectors across the field.  The
    filled circles represent the rotated rectilinear grid, and the
    vectors represent the offsets $\delta X$ and $\delta Y$ from this
    grid to the centroids of the grid mask spots; the vector lengths
    are magnified 5$\times$.  The subplots show the distribution
    of $\delta X$ and $\delta Y$ vs. $X$ and $Y$.  The bottom row
    displays the residual errors after applying the distortion
    correction, with vectors magnified 20$\times$.}
  \label{fig-distmap}}
\end{figure}

\new{
Given the acceptable repeatability of the H-50/50 + CH4-H4\%
distortion correction across all the instrument configurations, we
adopted it as the standard correction for all Campaign data.  The
distortion correction coefficients for this mode are listed
in Table~\ref{tab-dccoeffs}, and maps of the distortion correction
vectors and the post-correction residual errors are displayed in
Figure~\ref{fig-distmap}.}  More than half of
the Campaign data were taken in the dual-channel ASDI mode with this
configuration.  Applications which require the highest possible
accuracy, however, may wish to use more detailed calibration datasets
taken in a particular configuration.

\subsection{Instrument Alignment Angle}
\label{sec-iaa}
\nobreak
The Telescope Control System (TCS) uses a simple formula to control
the Cassegrain Rotator (CR) so that the desired position angle on sky,
{\ipa}, the {\em Instrument Position Angle}, is oriented
parallel to the detector columns toward the top of the image.  A 
quantity {\iaa}, the {\em Instrument Alignment
  Angle}, is defined to represent the rotation between
the CR and science detector reference frames.  The CR angle
\thcr\ is related to the other angles by
\begin{equation}
   \theta_{\rm CR} = \theta_{\rm IPA} - \theta_{\rm par} - \theta_{\rm IAA} + 180^\circ,
\end{equation}
where \thpar\ is the parallactic angle of the target under
observation.

For example, if a detector is mounted exactly aligned with the CR
frame, $\theta_{\rm IAA} = 0$.  To achieve $\theta_{\rm IPA} = 0$
(North up on detector) on the meridian where $\theta_{\rm par} = 0$,
\thcr\ would be 180$^\circ$.  Note that \iaa\ is dependent on the
mechanical orientation of the instrument on the telescope; small
rotations or other changes in the mounting may in turn require a
different \iaa\ setting in order to achieve an accurate \ipa.

Observations may be defined with CR in two modes: {\em Follow\/} mode, in
which case the TCS sets and continuously updates the CR to keep the
specified \ipa\ vertical on the detector, and {\em Fixed\/} mode where the TCS
sets the CR to a fixed value, allowing the sky (and therefore \ipa) to
rotate on the detector during the observation.  Follow mode is used
for most observations with other science instruments, but Fixed is
used for NICI ADI and ASDI observations.

For NICI, \iaa\ can represent only one of the two science
channels.  The natural choice is the Red channel, due to its smaller
DW-dependent systematic variations in scale and rotation than the
Blue's.  With this calibration, at $\theta_{\rm IPA} = 0$, north is exactly
vertical on the Red channel, while it is offset by $-1\fdg1$ on the
Blue channel.

Setting \iaa\ requires observing an astrometric standard
target or field, an exercise which must be performed each time NICI is
mounted on the telescope.  Table~\ref{tab-iaa} lists the individual
NICI mountings between 2008 and 2012, the astrometric standard
observed, and the \iaa\ and pixel scale results.  Over the course of the
Campaign, our calibration technique grew more sophisticated as we
added distortion corrections and switched from binary stars to a more
precise astrometric field.

\subsubsection{Binary Stars}
\label{sec-binstars}

For the Campaign's first two years, from 2008 August to 2010 October, we
used two binary stars for astrometric calibration.  An ideal binary
would have a separation between 5$''$ and 10$''$ (small enough to fit
into NICI's 18$''$ field yet large enough to provide a sufficient lever
arm for an accurate measurement), a separation precision $<$ a few mas,
and a position angle precision $< 0\fdg1$.  Such high precision requires
modern speckle measurements; orbits based on older historical data
are not sufficiently precise.  Several well-known systems with
precise orbits were rejected because their stars were too bright,
saturating on NICI even with the shortest possible exposure times.

We adopted \object{70 Oph} (STF~2272AB, WDS~18055+0230), which has an
accurate speckle-based orbit \citep{2000A&AS..145..215P,
  2008A&A...482..631E} and 5\farcs5 separation in mid-2008, as our
initial astrometric standard.  The $V = 4.2$ primary component was
bright enough to provide sufficient S/N when placed under the
coronagraph mask.  The off-mask image of the secondary component, with
$V = 6.2$, saturated slightly in the PSF core, but its position could
still be measured to $\pm 0.5$~pixel based on the unsaturated halo.
In this way, \iaa\ for mounting 1 was measured (see
Table~\ref{tab-iaa}), with an estimated uncertainty of $\pm 0\fdg09$.
When NICI changed ISS ports (from side-looking to up-looking) in late
2009, \object{70 Oph} was not accessible, so we observed the binary 
\objectname[CCDM~J22398]{HDO~171~B-C}
(CCDM~J22398-1942~B, WDS 22398-1942) instead.  \object{70 Oph} was
observed again for mounting 3 in 2010 March.  At this time we had not
developed the distortion-correction algorithm, so these data were
reduced and \iaa\ set with no distortion correction.

Follow-up checks and corrections of these early binary star
calibrations are described in \S\ref{sec-earlycheck}.

\subsubsection{Trapezium Cluster}
\label{sec-trapezium}
\nobreak At the start of mounting 4 in 2010 October, we observed the
Trapezium Cluster in Orion as an astrometric field.  We used the {\em
  HST}-derived coordinates from \citet{2008AJ....136.2136R}, applied
the grid mask distortion correction procedure described in
\S\ref{sec-distortion}, and fitted the measured position centroids to
the celestial coordinates as described for the LMC field in
\S\ref{sec-lmc}.  Although the technique appeared to be superior to
the binary star calibrations, the fitting errors were unacceptably
large: $\approx 2$~pixels rms, \new{considerably larger than
  what we eventually achieved for the LMC field using the same
  observing and reduction techniques.  The large errors
  may be caused by systematic errors in the
  {\em HST} optical astrometry or the underlying Two Micron All Sky
  Survey (2MASS) catalog \citep{2003tmc..book.....C} absolute astrometry against
  which the {\em HST} data were registered.}  The magnitude of these
errors led us to develop the LMC-based calibration described in the
next section.

Note that during this period in 2010 October-December, while new
software was being developed to apply the distortion correction,
measure the alignment, and compute \iaa\ changes, a number of
errors were made computing the direction of the \iaa\ 
corrections and the WCS parameters in the headers, as indicated in
Table~\ref{tab-iaa}.  Methods for correcting these errors are
described in \S\ref{sec-wcs}.

\subsubsection{LMC Astrometric Field}
\label{sec-lmc}
\nobreak
In 2010 November we began to observe a field in the Large Magellanic
Cloud (LMC) that was the subject of a detailed {\em HST}-based
astrometric calibration in support of {\em JWST}\/
\citep{2007ASPC..364...81D}.  For {\em JWST}, the field was chosen to
be relatively free of bright stars, but fortunately it contains three
$R = 11-12$~mag stars bright enough to serve as natural guide stars
for NICI's AO system.  The data file containing high-precision
coordinates was supplied to us by J.\,Anderson \citetext{private
communication}.

\begin{figure}[p]
\figurenum{3}
\epsscale{1.0}
\plotone{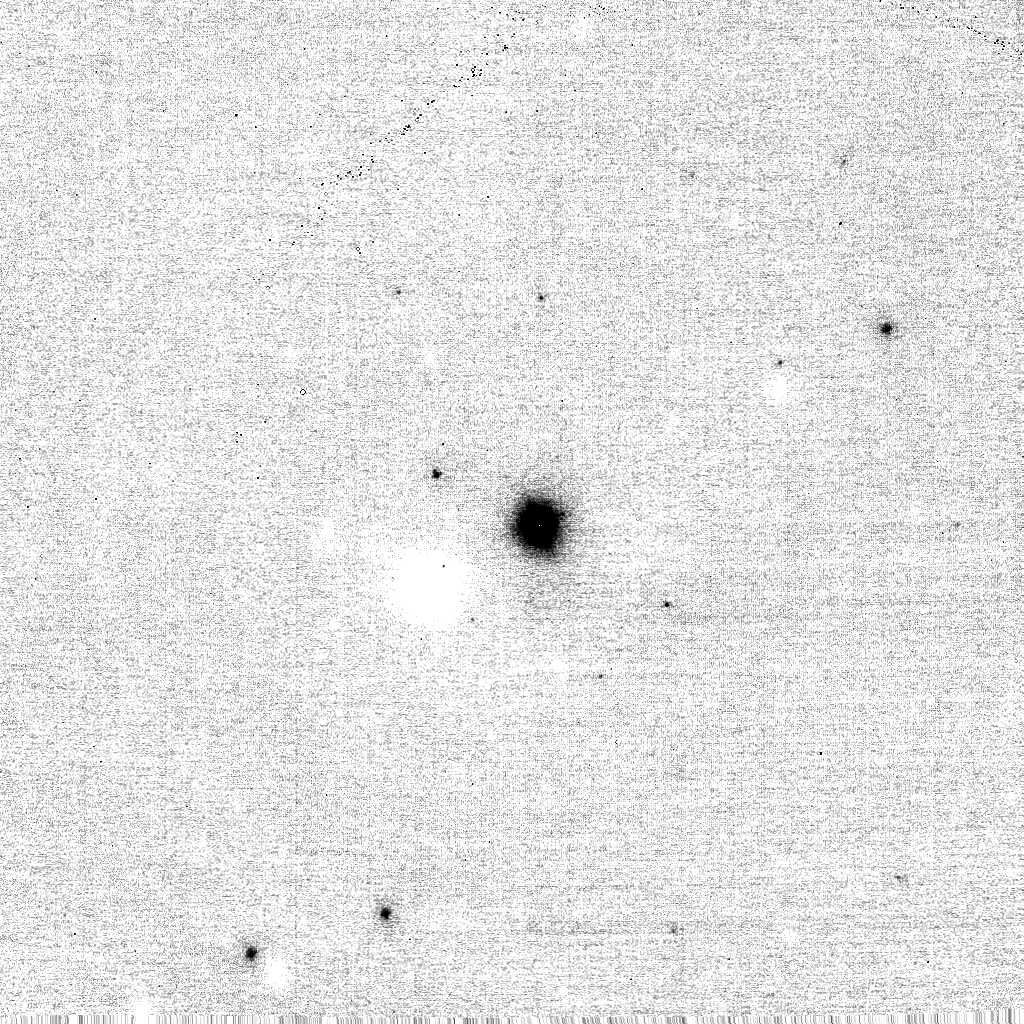}
\caption{LMC-11mag field in the NICI Red channel, CH4-H4\%L filter
  from 2012 December 22 UTC, with the $R \approx 11$~mag guide star
  at center.  This is a pair-subtracted image with a 4$''$ dither.
  \label{fig-lmc11mag}}
\end{figure}

The subfield with the brightest guide star, named ``f606w\_11.341'' in
the STScI documentation but which we refer to as ``LMC-11mag,''
contains 12 stars with $H < 17$ which can be detected by NICI
in about 5~minutes observing time.  For consistency, the field was
observed in Cass Rotator Follow mode at {\ipa}~=~0, always with
the Clear focal-plane mask (no coronagraph), the H-50/50 beamsplitter,
and the {CH4-H4\%L} and S filters.  Image pairs taken with a 4$''$
dither were subtracted to cancel the background emission
(Fig.~\ref{fig-lmc11mag}), then the standard distortion correction
was applied and the stars' centroid positions in the Red channel
measured.

The centroid position probable errors were estimated by synthesizing
images with Gaussian PSF's over a range of S/N combined with
backgrounds with random variations.  After the centroid for a given
S/N PSF was measured for each of the random backgrounds, the standard
deviation $\sigma$ could be computed directly, and a lookup table of
S/N vs. $\sigma$ constructed.  The probable error of each star in the
real image was then estimated by measuring its S/N and finding the
corresponding $\sigma$ in the lookup table.

The centroid positions were fitted with a $\chi^2$ minimization
technique similar to the grid mask fitting described by
Eqn.\ref{eq-chisq} to determine the pixel scale and the \iaa\ 
correction, using the probable errors $\sigma$ estimated for each
star.

Table~\ref{tab-lmc} displays the results of an LMC-11mag astrometric
fit taken for mounting~10 on 2012 December 22 UTC.  The $R = 11$~mag guide star
consistently shows a large 5 - 6 pixel error in all images, so it is
not included in the fit.  (The exact reason for the error is unknown;
the star is certainly a foreground star, but the error appears to be
too large to be explained by proper motion, and no proper motion was
detected between the different NICI astrometric images over a period
of 3 years.)  As with the pinhole-grid images, the value
of $\chi^2$ per degree of freedom $> 30$ indicates that the errors are
dominated by systematic effects.  In addition to the distortion
correction errors described in \S\ref{sec-fitproc}, additional sources
of error on a star field are errors in the provided celestial
coordinates, proper motions, atmospheric refraction, and
anisoplanicity effects which cause stars far from the center of the
field to become elongated.

The residual error between the expected and measured positions \new{of
  all the LMC field stars is 0.68~pixels rms, with no errors $\gtrsim
  1$~pixel, better than our goal of $< 1$ pixel across half the field.
  This 2012 December dataset was taken in good seeing and is one of
  our best results, but in all our calibrations the residual errors
  are $< 0.8$~pixels across the field, or $\approx 0.08$\% in scale
  and 0\fdg046 in rotation.  This corresponds to a center-to-edge
  accuracy of $< 0.4$~pixels or 7~mas, better than our 20~mas
  requirement.  The errors are similar for both up-looking and 
  side-looking port mountings, even with the extra reflection 
  on the side port.}

After 2010 the LMC-11mag field became the primary NICI astrometric
reference, observed regularly after most instrument mountings and port
changes and periodically during observing semesters.  We find that
\iaa\ varies by a few tenths degree between mountings, but
periodic checks within a mounting are consistent to $<0\fdg03$,
exceeding our calibration requirements.  We calibrated a secondary
field, \object{HIP 62403}, relative to LMC-11mag for use during the southern
winter months when the LMC is inaccessible; our derived astrometric
data for this field are listed in Table~\ref{tab-HIP62403}.

The LMC fits also yield the image pixel scale which, unlike {\iaa},
should not vary between mountings.  Over 12 separate LMC
measurements, we find a Red channel scale of
17.958$\pm$0.022~mas/pixel.  The relative probable error is 0.12\%, or
about 1 pixel across the detector, again within our requirements.

\subsubsection{Checks and Corrections of Early Calibrations}
\label{sec-earlycheck}
\nobreak
After developing the distortion correction and LMC-based calibration
in late 2010, we checked the earlier binary star calibrations
in two ways.  First, the binary star centroids were remeasured
after applying the grid mask distortion correction.  We found that the
mounting~1 and 2 calibrations were unaffected by the distortion
correction to a level of $\approx 0\fdg1$, because both binary components
were placed within the low-distortion central region.
For mounting~3, however, 70~Oph~B was placed in the lower left corner
of the field and suffered a distortion of $\approx 2.5$~pixels, causing a
rotation error of $0\fdg7$.

As a second check method, we utilized several Campaign target fields
with multiple background stars that had been observed both with binary
star and later LMC-based astrometric calibration.  The results are
summarized in Table~\ref{tab-astromcorr}.  This method verified the
accuracy of the mounting 1 and 2 calibrations and the error with
mounting 3.  Two good comparison fields for mounting 3 indicated an
error of $0\fdg5$ and a true {\iaa}~=~112.2, which we judge to be
more reliable than the 70~Oph derived value due to saturation of
70~Oph~B and its position in the corner of the field.  A detailed
procedure for correcting the \iaa\ error is presented in
\S\ref{sec-wcs}; these corrections were applied to the Campaign data
reductions, but {\em not\/} to the raw data stored in the Gemini
Science Archive.

These checks revealed one range of dates, 2009 April 26-27 UTC, for
which the field rotation is in error by $0\fdg5$, an unusually large
amount.  \object{Proxima Cen} and \object{HD 196544} data taken on
these dates both exhibit the anomaly.  We suspect that an error in the
Cass Rotator position or datum sometime between April 8 and 26 caused
this error.  After April 27, NICI operations were suspended for the
winter, so we do not have futher data until 2009 August, by which time
the Cass Rotator position is correct.

\subsection{WCS}
\label{sec-wcs}
\nobreak
The WCS in each NICI MEF extension specifies the mapping between sky
and detector for that channel, including the field rotation, pixel
scale, and reflection between sky and detector.  It has
limitations, however, in that it describes the absolute astrometric position
only to low accuracy and does not represent the image distortion.  In
addition, during NICI's operations there were many instances when \iaa\ 
was set incorrectly and therefore the WCS does not
represent the true astrometry.  Therefore, in this section we will
review how the WCS is constructed, its limitations and errors, and how
the errors can be corrected.

The Gemini algorithm that builds the WCS for each MEF file begins with
a mapping file that contains a series of $X, Y$ positions in detector
coordinates along with their corresponding on-sky angular offsets from
the field center in the telescope coordinate system.  The algorithm
fits a transformation to these points, then applies the current
telescope pointing and Cassegrain rotator angles to compute the on-sky
WCS.  The original intent was to populate the mapping file with points
measured by offsetting a star to different positions on the detector
and recording both the detector and celestial positions.  For NICI's
very narrow field of view and small pixels, however, this technique is
insufficiently precise. (It relies on the mapping accuracy of the
Peripheral Wavefront Sensor probe arm, which could contain errors of
several tenths of an arcsec.)  Therefore, we developed a simple script
to generate synthetic mapping files based on the pixel scale and \iaa\ 
rotation angle determined from the calibrations.  This
technique produces a WCS limited only by the quality of the
astrometric field measurements as described in preceeding sections.

The Red channel mapping file is generated directly from the measured
Red channel scale and {\iaa}.  The Blue channel file is generated
using the more precise scale and rotation angle relative to the Red
channel determined from the grid mask images.  For the LMC
calibrations after 2010 November, the Blue WCS rotation angle is set
$-1\fdg1$ from the Red angle, and the scale is 1/0.9980 = 1.0020 $\times$
the Red scale (with the scale, in mas/pixel, being inversely
proportional to the magnification).

Gemini's WCS standard does not include distortion parameters such as
the SIP system, which defines FITS header distortion coefficients to
supplement the WCS \citep{2005ASPC..347..491S}.  Therefore, for NICI
files after 2010 November, the WCS represents the coordinate system
{\em after}\/ the standard distortion correction (for H-50/50 +
CH4-H4\%) described in \S\ref{sec-distortion} has been applied.

In addition, note that the WCS represents the orientation at the {\em
  beginning}\/ of the exposure.  If the CR mode is Follow, then this
orientation is correct throughout the exposure.  However, if the CR
mode is Fixed, as is the case for ADI observations, a rotation must be
applied to correct the change in field angle between the start and
mid-point of the exposure.

\subsubsection{WCS  Corrections}
\label{sec-wcs-corr}
\nobreak
As we previously described in \S\ref{sec-binstars} and
\ref{sec-trapezium}, the \iaa\ and pixel scale values measured
during several calibrations contain errors.  These
errors are also manifested in the WCS.  Known issues are:
\begin{enumerate}
\item{The WCS values for each mounting are based on the pixel scale derived from
    a particular calibration dataset.  For highest
    accuracy and consistency, we recommend using the mean LMC-based scales of
    17.958~mas/pixel for the Red channel and 17.994 for Blue 
    for all epochs.}
\item{The values of \iaa\ and WCS may be incorrect for data taken
    immediately after each mounting, until astrometric calibrations
    could be taken and analyzed.  The affected dates are listed in Table~\ref{tab-iaa}.
    Mounting 4 is the most severe example: data from 2010 October
    and November contain several errors due to problems with the
    distortion-correction and astrometric fitting software used at
    that time for the Trapezium and LMC fields.}
\item{Before 2010-12-14, the rotation of the Blue WCS relative to the
    Red did not use the correct value of $-1\fdg1$.}
\item{For mounting 3 from 2010 March to October, the \iaa\ 
    value of 111.70 was too low by $0\fdg5$.}
\item{Data taken on 2009 April 26-27 have a rotation error of
    $0\fdg5$, apparently caused by a short-term error in the Cass Rotator.  (See last
    paragraph of \S\ref{sec-earlycheck}}
\end{enumerate}

The WCS rotation correction for the Red channel is listed in
Table~\ref{tab-iaa}.  In addition, we have developed an IDL script
that automatically corrects a MEF file's WCS according to the date
of observation (see Appendix).  Note that the script does not modify
the image data; only the WCS is updated to represent the true
astrometry of the data.  The updated file with the corrected WCS can
then be pipeline-processed as usual.

\section{Campaign Astrometry Statistics}
\label{sec-campstats}
\nobreak
Several dozen of the Campaign's target fields contained candidate
companions which required follow-up checks for common proper motion
and parallax.  The vast majority of these candidates were determined
to be background stars.  We can use these multi-epoch observations to
evaluate the consistency of the astrometric calibration over the
course of the Campaign.

Our technique is simply to compare the relative positions of the
primary target and background stars at the available epochs, after
correcting the primary's relative parallax and proper motion to the
first epoch.  Differences in the corrected positions between epochs
indicate errors in the position measurements themselves, the proper
motion correction, the astrometric calibrations, or a combination of
all three effects.  Some dense fields contained multiple background
stars, permitting more detailed checks independent of the proper
motion correction, but because many fields just contained the primary
and one background star we will discuss only the primary-to-background
position differences here.

\begin{figure}[p]
\figurenum{4}
\epsscale{1.1}
\plottwo{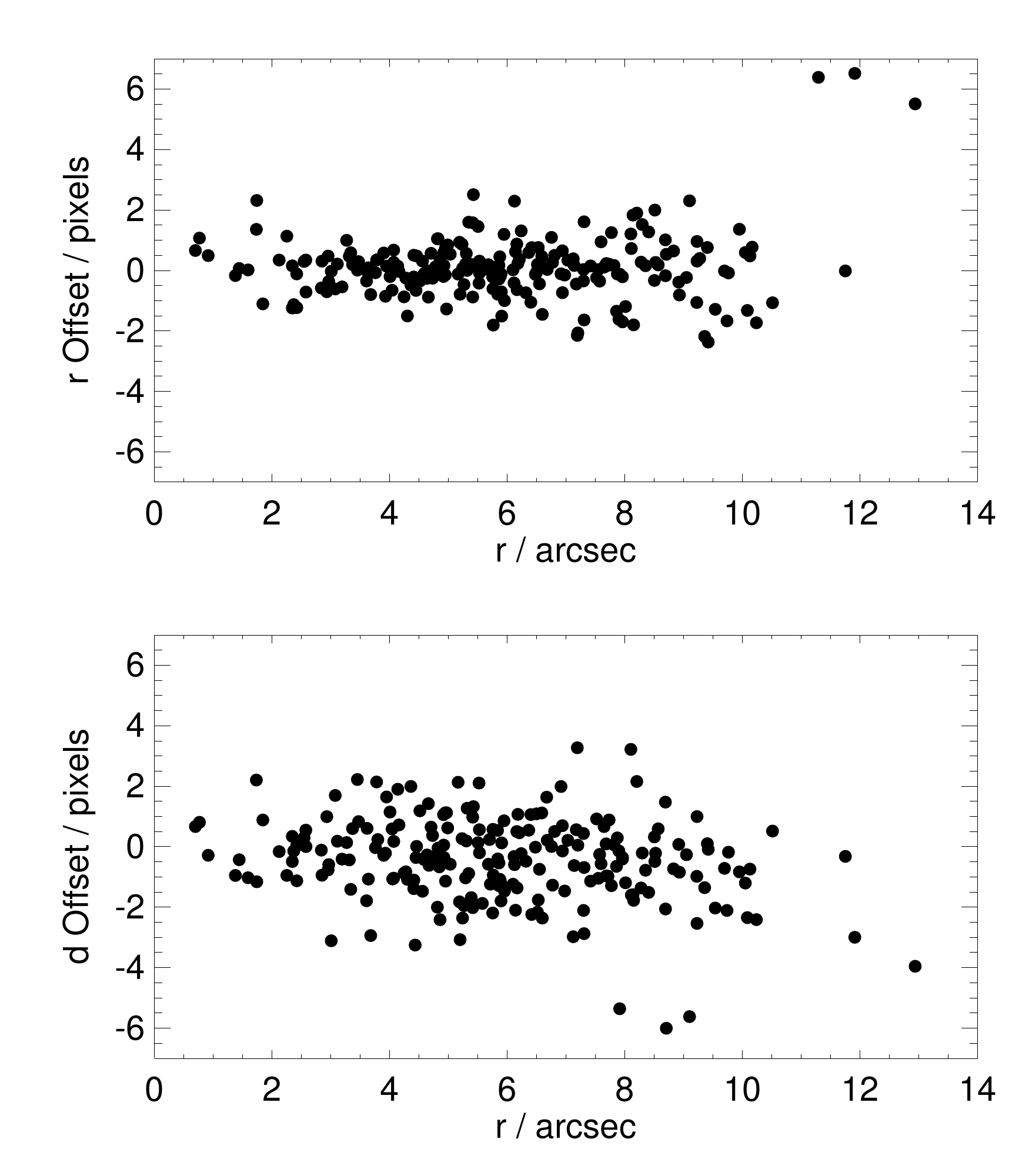}{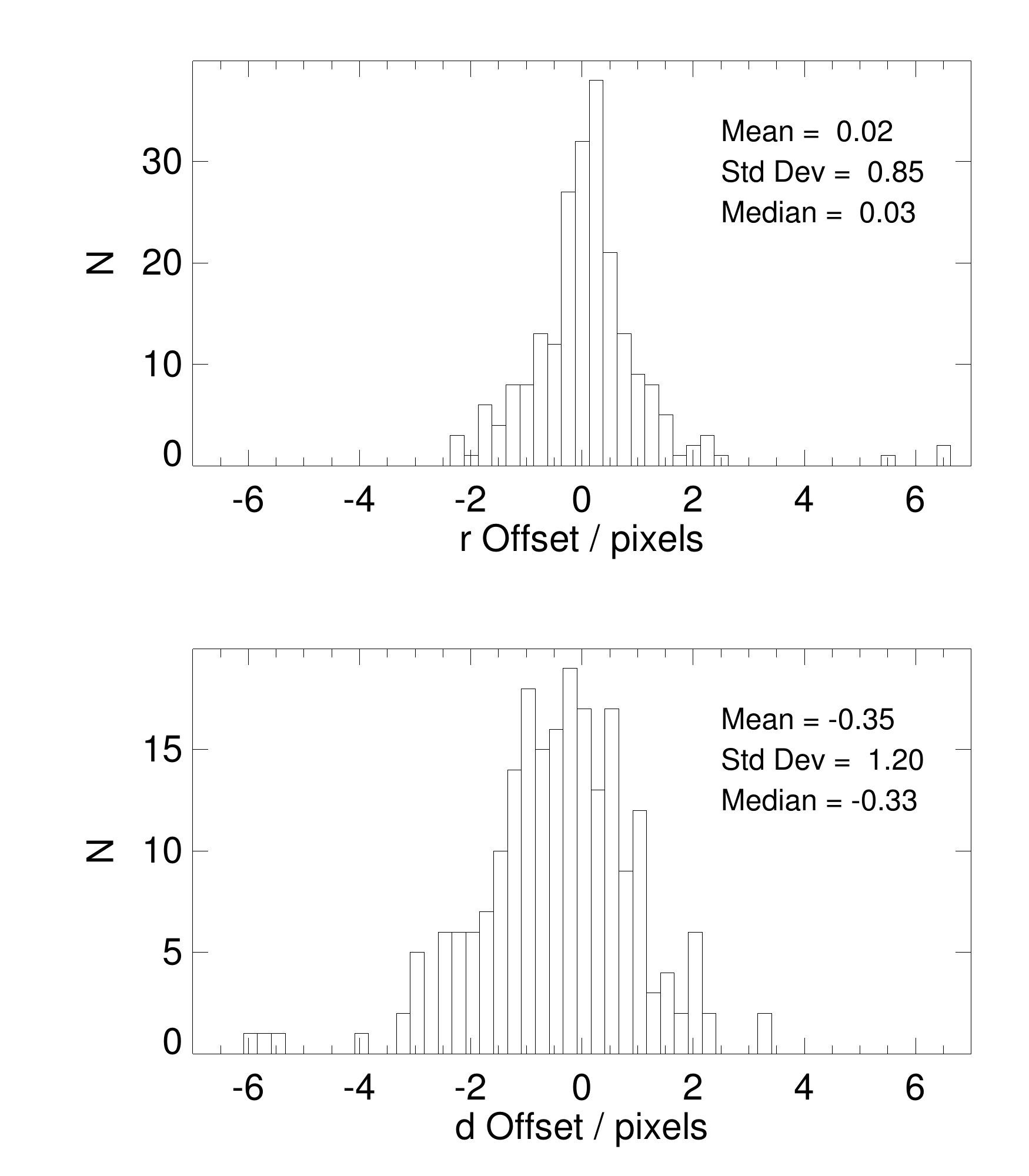}
\caption{Distribution of position offsets between two epochs, after
  correction of parallax and proper motion.  The quantity $r$ is the
  separation of the background star from the primary, and $d =
  r\,\tan(\theta)$ where $\theta$ is the position
  angle.\label{fig-offsetdistrib}}
\end{figure}

Figure~\ref{fig-offsetdistrib} shows the offset distributions for a
total of 218 pairs of observations of 205 background stars in 113
fields.  The offsets are indicated parallel and perpendicular to the
radius vector from the primary to the background star, which is
appropriate given the large derotations applied to the ADI frames
before stacking and helps to identify systematic errors in scale and
rotation.

The left panel of Figure~\ref{fig-offsetdistrib} shows the separation
$r$ and perpendicular $d = r\, \tan(\theta)$ offsets vs. $r$ itself.
The distribution of the $r$ offsets shows a possible increase with
radius, suggesting scale errors on the order of 0.1\%.  The $d$
offsets, however, maintain a consistent width from small to large $r$,
indicating that the primary source of error is in the individual
position measurements of the stars.  In other words, if the dominant
source of errors were in the rotation calibration, the distribution
would be expected to increase proportional to $r$, which is not
observed.

The right panel shows the distribution histograms in the same pixel
units.  The distributions have a central core with $\sigma =
0.85$~pixels in $r$ and 1.20 in $d$; values that are only slightly
larger than the rms errors of the distortion correction and the LMC
field astrometric fits.  Above 3.5~pixels total offset lie 9 {\it
  outlier} points, all of which have $r > 7''$.  The large errors in
these cases are caused by inaccurate distortion correction near the
edges of the field and very different positions of the stars on the detector
between the two epochs -- due to either different field angles, or to
the two observations being taken on different telescope ports with an
odd and even number of reflections, which reversed the image and
caused the same object to appear on opposite sides of the detector.

That the primary source of the $d$ offset errors appears to be due to
individual position errors does not have an obvious explanation.
Because the PSF core FWHM is usually 3-4~pixels, centroid measurement
errors as large as 3~pixels would typically be several $\sigma$.  It
should be noted that the astrometry is measured from final images
which are coadded from many derotated ADI frames, which tends to
broaden the individual stars.  Possible causes of systematic error are
small-scale irregularities in the distortion correction, shifts in the
position of the primary star which is dimmed by more than 6 magnitudes
by the partially transparent occulting mask, or other irregularities
in the mask.

\section{Conclusions}
\label{sec-conclusions}
\nobreak \new{We have developed an astrometric calibration for the
  NICI Planet-Finding Campaign based on a grid mask distortion
  correction and an LMC astrometric field.  The accuracy
  achieved by the calibration is $\lesssim0.08$\% in scale and
  $\lesssim0\fdg046$ in rotation, corresponding to $\lesssim7$~mas
  center-to-edge.}  The calibration for each channel is represented by
the MEF header WCS.  Before 2010 November the calibrations and WCS
contain known errors; we provide tables and an IDL routine to correct
archival NICI data to the final calibration.

\acknowledgements

This work was supported in part by NSF grants AST-0713881 and
AST-0709484.  The authors thank the NICI team members J.~Hinds and
C.~Lockhart, and Gemini engineers R.~Galvez, G.~Gausachs, J.~Luhrs,
G.~Perez, R.~Rogers, R.~Rojas, L.~Solis, G.~Trancho, and C.~Urrutia
for their expert support of NICI during commissioning and operations.
F.~Rantakyro, E.~Christensen, and E.~Artigau provided valuable
operational and software support.  We thank J. Anderson for supplying
the {\em HST\/} data for the LMC fields, and the anonymous referee
for providing valuable suggestions.  This research has made use of
the SIMBAD database, operated at CDS, Strasbourg, France; the
Washington Double Star Catalog maintained at the U.S. Naval
Observatory; and NASA's Astrophysics Data System.

{\it Facilities:} \facility{Gemini (NICI)}

\appendix

\section{NICI MEF Format}
\label{app-mefformat}
The two NICI science channels are Red (transmitted light through DW)
and Blue (reflected by DW).  The Red and Blue channels are also
referred to as ``Holmes'' and ``Watson,'' respectively.

The MEF files contain a primary header unit (PHU) followed by two extensions:
\medskip
\begin{enumerate}
   \item Primary Header 
   \item Extension 1: Red  (holmes) channel header and data
   \item Extension 2: Blue (watson) channel header and data
\end{enumerate}

\bigskip

Selected astrometry-related keywords in an example NICI MEF Header.

\footnotesize
\begin{verbatim}

# PRIMARY HEADER UNIT
OBJECT  = 'f606w_11.341'       / Object Name
FRAME   = 'FK5     '           / Target coordinate system
EQUINOX =                2000. / Equinox of coordinate system
EPOCH   =                2000. / Target Coordinate Epoch
RA      =          80.48629583 / Target Right Ascension
DEC     =         -69.44843889 / Target Declination
HA      = '-00:52:22.05'       / Telescope hour angle
PA      =                   0. / Sky Position Angle at start of exposure (IPA)
INPORT  =                    1 / Number of ISS port where instrument is located 
CRPA    =    -49.5685362028393 / Current Cass Rotator Position Angle
CRMODE  = 'FOLLOW  '           / Cass Rotator Mode
IAA     =                247.5 / Instrument Alignment Angle
FOCS    = 'Open    '           / Fiber Optic Calib. Src.
FPMW    = 'Clear_G5710'        / NICI Focal Plane Mask Wheel
DICHROIC= 'H-50/50_G5701'      / NICI Dichroic Wheel

# EXTENSION 1 HEADER
CHANNEL = 'RED     '           / NICI Science Camera Channel
FILTER_R= 'CH4-H4\%L_G0740'    / NICI Red Filter Wheel                               
RADECSYS= 'FK5     '           / R.A./DEC. coordinate system reference          
CTYPE1  = 'RA---TAN'           / R.A. in tangent plane projection               
CTYPE2  = 'DEC--TAN'           / DEC. in tangent plane projection               
CRVAL1  =           80.4883844 / RA at Ref pix in decimal degrees               
CRVAL2  =          -69.4482734 / DEC at Ref pix in decimal degrees              
CRPIX1  =           413.999522 / Ref pix of axis 1                              
CRPIX2  =           594.997314 / Ref pix of axis 2                              
CD1\_1  =       5.01646534E-06 / WCS matrix element 1 1                         
CD1\_2  =       8.26096075E-11 / WCS matrix element 1 2                         
CD2\_1  =      -3.29874171E-10 / WCS matrix element 2 1                         
CD2\_2  =      -5.01703714E-06 / WCS matrix element 2 2

# EXTENSION 2 HEADER
CHANNEL = 'BLUE    '           / NICI Science Camera Channel                         
FILTER_B= 'CH4-H4\%S_G0743'    / NICI Blue Filter Wheel                              
RADECSYS= 'FK5     '           / R.A./DEC. coordinate system reference          
CTYPE1  = 'RA---TAN'           / R.A. in tangent plane projection               
CTYPE2  = 'DEC--TAN'           / DEC. in tangent plane projection               
CRVAL1  =           80.4883844 / RA at Ref pix in decimal degrees               
CRVAL2  =          -69.4482734 / DEC at Ref pix in decimal degrees              
CRPIX1  =           570.000515 / Ref pix of axis 1                              
CRPIX2  =           650.997321 / Ref pix of axis 2                              
CD1\_1  =       -5.0159887E-06 / WCS matrix element 1 1                         
CD1\_2  =      -6.99793144E-08 / WCS matrix element 1 2                         
CD2\_1  =       7.03997325E-08 / WCS matrix element 2 1                         
CD2\_2  =      -5.01655469E-06 / WCS matrix element 2 2

\end{verbatim}
\normalsize

\bigskip

\section{WCS Corrections}
\label{app-wcscorr}
An IDL routine {\tt NICI\_FIXWCS} is available at 
\url{http://www.gemini.edu/sciops/instruments/nici/data-format-and-reduction}
to correct known errors in the
NICI WCS.  It automatically applies corrections stored in lookup
tables according to the observation date, using the following
algorithm:

\begin{enumerate}
\item Extract the ISS Port ({\tt INPORT}) and $\theta_{\rm IAA,hdr}$ ({\tt IAA}) from the Primary Header Unit and the Red WCS from extension 1.
\item Compute the WCS rotation angle $\theta_{\rm hdr}$.
\item Determine $\theta_{\rm IAA,true}$ from a lookup table according to the observation date.
\item Compute $\theta_{\rm corr} = \theta_{\rm IAA,true} - \theta_{\rm IAA,hdr}$.
\item Compute the new Red  WCS angle $\theta_{\rm red}  = \theta_{\rm hdr} + \theta_{\rm corr}$.
\item Compute the new Blue WCS angle $\theta_{\rm blue} = \theta_{\rm red} - 1\fdg1$.
\item Compute the new Red and Blue WCS's with a left-handed coordinate
  system using the mean pixel scales and new rotation angles.
\item Reflect the Red WCS left-right for the side-looking port or the
  Blue WCS left-right for the up-looking port.
\item Write the new WCS's into the MEF extension headers.
\end{enumerate}

These WCS computations can also be performed with standard WCS processing
routines such as the IDL Astronomy Library astrometry routines
{\tt GETROT} and {\tt HROTATE}.  Table~\ref{tab-wcscorr} lists the
date-specific $\theta_{\rm IAA,true}$ values used by {\tt NICI\_FIXWCS}.

A typical data-reduction strategy is outlined below.  Additional
details, up-to-date information, and the distortion correction files
are available at
\url{http://www.gemini.edu/sciops/instruments/nici/data-format-and-reduction}.

\begin{enumerate}
   \item Apply recommended corrections to the Red and Blue channel WCS's.
   \item Apply the standard distortion corrections to the Red and Blue data.
   \item For CassRot Fixed data, apply a rotation to the WCS to
     correct it to the mid-point of the exposure. The
     amount of correction is proportional to the rate of change of the
     parallactic angle during the exposure, which may vary
     significantly as targets rise, transit, and set.  At this point
     the WCS should accurately represent the astrometric properties
     of the data.
   \item Reflect both channels up-down so that pixel (1,1) appears in
     the lower-left corner of standard image display tools, using a routines such
     the IDL Astronomy Library's {\tt HROTATE} that also reflects
     the WCS.
   \item Depending on the ISS port, reflect either the Red or Blue
     channel left-right to produce a left-handed coordinate system.
   \item Apply further processing (ADI pipeline, derotation, etc.)
     using the WCS to correct the rotation and scale of individual
     exposures.
\end{enumerate}

\clearpage


\clearpage

\begin{deluxetable}{lccccccccccc}
\tablecolumns{12}
\tablewidth{0pc}
\tabletypesize{\footnotesize}
\setlength{\tabcolsep}{4pt}  
\tablecaption{Distortion Correction Parameters\label{tab-dcparams}}
\tablehead{
\colhead{} & \multicolumn{4}{c}{Red Channel}& \colhead{} & \multicolumn{4}{c}{Blue Channel} & \multicolumn{1}{c}{ } \\ 
\cline{2-5} \cline{7-10} \\
\colhead{DW}       & 
\colhead{Filter}   & \colhead{$d$}       & \colhead{$\theta$}   & \colhead{rms Err}  & &
\colhead{Filter}   & \colhead{$d$}       & \colhead{$\theta$}   & \colhead{rms Err}  & 
\colhead{B/R Mag}  & \colhead{R-B Rot}  \\ 
\colhead{}         & 
\colhead{}         & \colhead{(pixels)}  & \colhead{($^\circ$)} & \colhead{(pixels)} & &
                   & \colhead{(pixels)}  & \colhead{($^\circ$)} & \colhead{(pixels)} & 
\colhead{}         & \colhead{($^\circ$)}}
\startdata
H-50/50 (max)\tablenotemark{a}& 4\%L & 29.574 & 22.313 &0.54 & & 4\%S & 29.514 & -21.210 & 0.53 & 0.9980 & -1.103   \\ 
H-50/50 (min)                 & 4\%L & 29.574 & 22.311 &0.54 & & 4\%S & 29.507 & -21.148 & 0.53 & 0.9977 & -1.163   \\ 
Mirror                        &  --  &   --   &   --   & --  & & 4\%S & 29.495 & -21.125 & 0.56 &  --    &   --     \\ 
Open                          & 4\%L & 29.568 & 22.382 &0.57 & &  --  &   --   &    --   &  --  &  --    &   --     \\
H/K                           & 4\%L & 29.560 & 22.286 &0.57 & & 4\%S & 29.487 & -21.213 & 1.01 & 0.9974 & -1.073   \\ 
H/K                           &  Ks  & 29.562 & 22.293 &0.57 & &  H   & 29.484 & -21.206 & 0.59 & 0.9974 & -1.087
\enddata
\tablecomments{Derived from grid mask data taken 2010 Dec 14 (H-50/50, Mirror, Open) and 2011 May 12 (H/K). 
The filter names 4\%L and 4\%S refer to the CH4-H4\%L and S filters, respectively, $d$ and $\theta$ are 
the separation and rotation of the grid spots for each channel, and rms Err is the rms error of the fit.
The last two columns indicate the magnification and rotation of the Blue channel relative to the Red.}
\tablenotetext{a}{The terms ``max'' and ``min'' indicate extremes in the range of the 
Blue channel spot positions due to the
non-repeatable DW detent for H-50/50, as explained in the text.}
\end{deluxetable}

\clearpage

\begin{deluxetable}{lccccccc}
\tablecolumns{8}
\tablewidth{0pc}
\tabletypesize{\footnotesize}
\tablecaption{Reference Spot Positions \label{tab-spotpos}}
\tablehead{
 \colhead{}       & \multicolumn{3}{c}{Red Channel}  & \colhead{}    & \multicolumn{3}{c}{Blue Channel}  \\ 
 \cline{2-4}  \cline{6-8} \\
 \colhead{DW}     & \colhead{Filter} & \colhead{$X$} & \colhead{$Y$} & \colhead{} & 
 \colhead{Filter} & \colhead{$X$}    & \colhead{$Y$} }
\startdata
H-50/50 (max) & 4\%L & 530 & 514 &  & 4\%S & 450 & 566 \\
H-50/50 (min) & 4\%L & 530 & 514 &  & 4\%S & 451 & 542 \\
Mirror        &  --  & --  & --  &  & 4\%S & 455 & 580 \\
Open          & 4\%L & 529 & 513 &  &  --  &  -- &  -- \\
H/K           & 4\%L & 531 & 508 &  & 4\%S & 465 & 593 \\ 
H/K           &  Ks  & 531 & 508 &  &  H   & 465 & 593 
\enddata
\tablecomments{The reference spot (same physical pinhole in the grid mask) positions in pixels for the datasets 
  in Table~\ref{tab-dcparams}.  The filter names 4\%L and 4\%S refer to the CH4-H4\%L and S filters, respectively. }
\end{deluxetable}

\begin{deluxetable}{lccccc}
\tablecolumns{6}
\tablewidth{0pc}
\tabletypesize{\footnotesize}
\tablecaption{Distortion Correction Residuals \label{tab-dch5050}}
\tablehead{
\multicolumn{1}{ c }{ } & \multicolumn{2}{ c }{Red Channel}      &  & \multicolumn{2}{ c }{Blue Channel} \\ 
\cline{2-3} \cline{5-6} \\
\colhead{DW}            & \colhead{Filter} & \colhead{rms Error} &  & \colhead{Filter} & \colhead{rms Error} \\
                        &                  &   (pixels)          &  &                  &  (pixels) }
\startdata
H-50/50 & 4\%L & 0.54 & & 4\%S & 0.53 \\ 
Mirror  &  --  &      & & 4\%S & 0.66 \\ 
Open    & 4\%L & 0.62 & &  --  & --   \\ 
H/K     & 4\%L & 0.62 & & 4\%S & 0.67 \\ 
H/K     & Ks   & 0.68 & &  H   & 0.72
\enddata
\tablecomments{Residuals after applying the distortion correction derived for the H-50/50, CH4-H4\% configuration 
to itself and four other configurations.  The residuals for the other four cases are only slightly greater
than the first, indicating that the same distortion correction may be applied to all configurations.}
\end{deluxetable}

\clearpage

\begin{deluxetable}{rrrrrrr}
\tablecolumns{7}
\tablewidth{0pc}
\tabletypesize{\footnotesize}
\setlength{\tabcolsep}{4pt}  
\tablecaption{\new{Distortion Correction Coefficients: H-50/50 BeamSplitter} \label{tab-dccoeffs}}
\tablehead{\colhead{} & \multicolumn{1}{c}{i = 0} & \multicolumn{1}{c}{i = 1} & \multicolumn{1}{c}{i = 2} & 
\multicolumn{1}{c}{i = 3} & \multicolumn{1}{c}{i = 4} & \multicolumn{1}{c}{i = 5}}
\startdata
\sidehead{Red channel $P_{i,j}$, CH4-H4\%L filter}
j=0 & 3.30330e+00 &-3.24809e-02 & 1.71647e-04 &-4.15214e-07 & 4.42279e-10 &-1.64635e-13 \\
1 & 9.56919e-01 & 6.51225e-04 &-4.02601e-06 & 1.04369e-08 &-1.13916e-11 & 4.35036e-15 \\
2 & 2.56604e-04 &-3.82735e-06 & 2.39133e-08 &-6.28410e-11 & 6.92881e-14 &-2.67375e-17 \\
3 &-6.51553e-07 & 9.23475e-09 &-5.73017e-11 & 1.50925e-13 &-1.66960e-16 & 6.47106e-20 \\
4 & 7.14946e-10 &-9.69040e-12 & 5.96375e-14 &-1.56947e-16 & 1.73795e-19 &-6.75174e-23 \\
5 &-2.79985e-13 & 3.67357e-15 &-2.24523e-17 & 5.89872e-20 &-6.53292e-23 & 2.54179e-26 \\
\sidehead{Red channel $Q_{i,j}$, CH4-H4\%L filter}
j=0 & 1.42582e+01 & 9.41488e-01 & 2.18931e-04 &-5.22009e-07 & 6.23837e-10 &-2.70063e-13 \\
1 &-5.40761e-02 & 6.07982e-04 &-3.77042e-06 & 9.95836e-09 &-1.15149e-11 & 4.81094e-15 \\
2 & 0.000191206 &-3.30812e-06 & 2.08871e-08 &-5.49720e-11 & 6.29836e-14 &-2.60687e-17 \\
3 &-3.77292e-07 & 7.52945e-09 &-4.79591e-11 & 1.26167e-13 &-1.44296e-16 & 5.96523e-20 \\
4 & 3.80158e-10 &-7.67197e-12 & 4.89736e-14 &-1.28818e-16 & 1.47450e-19 &-6.10716e-23 \\
5 &-1.42414e-13 & 2.88797e-15 &-1.84127e-17 & 4.84259e-20 &-5.55290e-23 & 2.30676e-26 \\
\sidehead{Blue channel $P_{i,j}$, CH4-H4\%S filter}
j=0 &-1.49018e+00 & 1.07896e-03 & 1.57155e-05 &-4.13557e-08 & 3.27833e-11 &-8.19736e-15 \\
1 & 9.93247e-01 & 6.31284e-05 &-4.91790e-07 & 1.18236e-09 &-1.03838e-12 & 2.99391e-16 \\
2 & 8.41072e-05 &-5.02536e-07 & 3.61158e-09 &-8.97672e-12 & 8.16102e-15 &-2.40618e-18 \\
3 &-2.42011e-07 & 1.44899e-09 &-1.03877e-11 & 2.65441e-14 &-2.48127e-17 & 7.50551e-21 \\
4 & 2.74100e-10 &-1.79337e-12 & 1.29831e-14 &-3.36450e-17 & 3.20101e-20 &-9.88390e-24 \\
5 &-1.10542e-13 & 8.03376e-16 &-5.82388e-18 & 1.51399e-20 &-1.45466e-23 & 4.55593e-27 \\
\sidehead{Blue channel $Q_{i,j}$, CH4-H4\%S filter}
j=0 & 1.17133e+01 & 9.88950e-01 &-5.97344e-05 & 1.98457e-07 &-2.06889e-10 & 7.90131e-14 \\
1 &-6.68061e-03 &-3.01784e-04 & 1.63966e-06 &-3.76719e-09 & 3.90648e-12 &-1.50652e-15 \\
2 &-6.09404e-05 & 1.72898e-06 &-9.59428e-09 & 2.21625e-11 &-2.30184e-14 & 8.88496e-18 \\
3 & 1.92937e-07 &-4.21968e-09 & 2.39915e-11 &-5.58094e-14 & 5.78653e-17 &-2.22094e-20 \\
4 &-2.04690e-10 & 4.55594e-12 &-2.63308e-14 & 6.15381e-17 &-6.36298e-20 & 2.42601e-23 \\
5 & 7.95641e-14 &-1.77333e-15 & 1.03499e-17 &-2.42680e-20 & 2.50371e-23 &-9.49402e-27
\enddata
\end{deluxetable}

\clearpage

\begin{deluxetable}{rlcclcccccl}
\tablecolumns{11}
\tablewidth{0pc}
\tabletypesize{\scriptsize}
\setlength{\tabcolsep}{4pt}  
\tablecaption{NICI Mountings and Astrometric Calibrations\label{tab-iaa}}
\tablehead{
 & & & & & & \multicolumn{4}{c}{WCS} \\
 \cline{7-10} \\
 \colhead{}                      & \colhead{Mounting}             & 
 \colhead{Port}                  & \colhead{WCS Revision}         & 
 \colhead{Target}                & \colhead{$\theta_{\rm IAA,orig}$} & 
 \multicolumn{2}{c}{Scale}       & 
 \colhead{$\theta_{\rm R-B}$}    & \colhead{$\theta_{\rm corr}$}  \\
 \colhead{\#}                    & \colhead{Date UTC}  &
 \colhead{}                      & \colhead{Date UTC}  & 
 \colhead{}                      & \colhead{}          &
 Red                             & Blue                &
                                 & Red                 &
 \colhead{Example Dataset}       \\
 \colhead{}                      & \colhead{}          &
 \colhead{}                      & \colhead{}          & 
 \colhead{}                      & \colhead{($^\circ$)}& 
 \multicolumn{2}{c}{(mas/pixel)} &
 \colhead{($^\circ$)}            & \colhead{($^\circ$)}
}
\startdata
 1&2008-07-27&5&10-07&70 Oph     &112.30                       &17.900&17.900&$-0.8$\phn\phn&   --                   &GS-NICI-COMM1-318       \\ 
 2&2009-10-24&1&10-29&HDO 171 B-C&247.50                       &18.060&18.060&              &   --                   &Eng files 20101024 1-11 \\ 
 3&2010-03-20&5&05-09&70 Oph     &{\bf 111.70}\tablenotemark{a}&17.800&17.800&              &$+0.50$&GS-CAL20100509-2                         \\ 
 4&2010-10-20&5&10-28&Trapezium  &{\bf 111.32}\tablenotemark{b}&17.910&17.910&              &$+1.29$&GS-CAL20101031                           \\ 
  &          &5&11-02&Trapezium  &{\bf 112.20}\tablenotemark{c}&17.960&17.960&              &$+1.29$&                                         \\
  &          &5&11-10&           &                             &      &      &              &$+0.41$\tablenotemark{d}&                        \\
  &          &5&11-20&           &                             &      &      &$-$1.137      &$+0.41$                 &                        \\
  &          &5&12-14&LMC-11mag  &112.61\tablenotemark{e}      &17.932&17.970&$-$1.1\phn\phn&$+0.0$\tablenotemark{e} &GS-CAL20101225-2        \\ 
 5&2011-01-14&1&02-16&LMC-11mag  &247.50                       &17.973&18.009&              &   --                   &GS-CAL20110117-2        \\ 
 6&2011-03-11&1&03-14&LMC-11mag  &247.50                       &17.973&18.009&              &   --                   &GS-CAL20110314-2        \\ 
 7&2011-04-27&1&05-12&LMC-11mag  &247.50                       &17.973&18.009&              &   --                   &GS-CAL20110512-1        \\ 
 8&2011-06-08&5&06-24&HIP 62403  &112.82                       &17.978&18.014&              &   --                   &GS-CAL20110626-3        \\ 
 9&2012-02-15&5&02-21&LMC-11mag  &112.64                       &17.950&17.986&              &   --                   &GS-CAL20120331-1        \\ 
10&2012-07-13&5&08-27&LMC-11mag  &112.42                       &17.936&17.973&              &   --                   &GS-CAL20121222-5        \\
11&2013-05-07&5&05-12&HIP 62403  &112.25                       &17.936&17.973&              &   --                   &GS-CAL20130512-1        \\
12&2013-06-19&1&06-21&HIP 62403  &247.50                       &17.936&17.973&              &   --                   &GS-CAL20130621-1   
\enddata
\tablecomments{Columns are: \\ 
\#: Number of NICI Mounting on the telescope, requiring a new \iaa\ measurement. \\
Port: ISS port; 5 = side-looking, 1 = up-looking. \\
WCS Revision Date: Date of \iaa\ and WCS verification or revision for the Mounting. \\
Target: Star or Field observed for astrometric measurements. \\
$\theta_{\rm IAA,orig}$: Originally derived \iaa\ value.  Values found to be erroneous upon later checks indicated in {\bf boldface}; see \S\ref{sec-earlycheck}.\\
Scale: Red and Blue scales used to construct the header WCS.\\
$\theta_{\rm R-B}$: Rotation between Red and Blue header WCS's; the true rotation is always $-1\fdg1$.\\
$\theta_{\rm corr}$: Angle by which to rotate Red WCS to correct calibration error; see Appendix \ref{app-wcscorr}.\\
Example Dataset: A representative astrometric data set taken after the final WCS update.
}
\tablenotetext{a}{Header value of \iaa\ erroneous due to image distortion.}
\tablenotetext{b}{Trapezium-based \iaa\ correction applied with wrong sign.}
\tablenotetext{c}{Corrected {\iaa}, but introduced Red channel WCS Error.}
\tablenotetext{d}{Corrected Red channel WCS Error.}
\tablenotetext{e}{Final LMC-11mag-based Mounting 4 calibration with updated $\theta_{\rm R-B}$.}
\end{deluxetable}

\clearpage

\begin{deluxetable}{rrccccccccccccccc}
\tablecolumns{17}
\tablewidth{0pc}
\tabletypesize{\scriptsize}
\tablecaption{LMC-11mag Field Astrometric Data, 2012-12-22 UTC \label{tab-lmc}}
\tablehead{
  \colhead{}                   & \colhead{}                   & 
    \multicolumn{4}{c}{Coordinates} & & & & &
    \multicolumn{3}{c}{Meas'd from Image} & &  
    \multicolumn{3}{c}{Transformed} \\
  \cline{3-6} \cline{11-13} \cline{15-17} \\
  \colhead{\#}                 & \colhead{Name}               & 
  \colhead{$\alpha$ (J2000.0)} & \colhead{$\delta$ (J2000.0)} & 
  \colhead{$X_0$}              & \colhead{$Y_0$}              & 
  \colhead{}                   & \colhead{S/N}                & 
  \colhead{$\sigma$}           & \colhead{}                   & 
  \colhead{$X_i$}              & \colhead{$Y_i$}              & 
  \colhead{$\Delta$}           & \colhead{}                   & 
  \colhead{$X_{t,i}$}          & \colhead{$Y_{t,i}$}          & 
  \colhead{$\Delta$}
}
\startdata
 0 & 46787\tablenotemark{a} &80.486295&-69.448440 & 538.2 &495.8& & 3500 & 0.01 & & 535.7 &499.7& 4.58 & & 535.7& 499.7& 4.58 \\
 1 & 46775 &80.490369&-69.450562 & 251.4 & 70.4 & & 71 & 0.09 & & 251.3 & 70.5& 0.19 & & 251.2&  70.6& 0.28 \\
 2 & 46767 &80.488462&-69.450358 & 385.7 &111.3 & & 86 & 0.08 & & 385.7 &110.3& 0.99 & & 385.6& 110.3& 0.96 \\
 3 &  3014 &80.488284&-69.447263 & 398.2 &731.8 & & 15 & 0.28 & & 397.8 &732.7& 0.95 & & 397.8& 732.7& 0.95 \\
 4 & 46799 &80.487740&-69.448175 & 436.5 &549.0 & & 82 & 0.08 & & 436.7 &549.6& 0.67 & & 436.7& 549.6& 0.69 \\
 5 &  3421 &80.487645&-69.448639 & 443.1 &457.7 & & 14 & 0.29 & & 443.5 &457.5& 0.43 & & 443.5& 457.5& 0.42 \\
 6 &  2771 &80.486258&-69.447291 & 540.8 &726.2 & & 23 & 0.20 & & 541.4 &726.3& 0.64 & & 541.4& 726.3& 0.68 \\
 7 & 46768 &80.484459&-69.448820 & 667.4 &419.7 & & 34 & 0.16 & & 667.0 &418.9& 0.84 & & 667.0& 418.9& 0.87 \\
 8 &  3694 &80.484371&-69.450446 & 673.6 & 93.6 & & 14 & 0.29 & & 674.2 & 94.4& 0.99 & & 674.1&  94.4& 0.92 \\
 9 &  2518 &80.482859&-69.447614 & 780.0 &661.4 & & 19 & 0.23 & & 779.7 &661.1& 0.47 & & 779.7& 661.1& 0.48 \\
10 &  2053 &80.481952&-69.446614 & 843.9 &862.3 & & 12 & 0.32 & & 843.5 &862.1& 0.44 & & 843.6& 862.0& 0.42 \\
11 & 46772 &80.481336&-69.447447 & 887.2 &694.9 & & 88 & 0.08 & & 886.9 &695.1& 0.36 & & 886.9& 695.0& 0.30 \\
   & & &\multicolumn{6}{r}{$\chi^2 = 348$, $\chi^2/\mathrm{DOF} = 34.8$} &\multicolumn{3}{r}{rms $\Delta$ = } & 0.69 & & & & 0.68
\enddata
\tablecomments{Example astrometric data for Mounting 10, side-looking Port, for the Red channel with the H-50/50 beamsplitter
and CH4-H4\%L filter.  For each star, the $\alpha$, $\delta$ coordinates from {\em HST}\/ are indicated in degrees; 
$X_0$ and $Y_0$ are the corresponding pixel coordinates after applying the nominal WCS transformation: scale = 17.95 mas/pixel, 
rotation = 0; $X_i$ and $Y_i$ are the measured image coordinates; and $X_{t,i}$ and
$Y_{t,i}$ are the transformed image coordinates after fitting, with scale 17.9553 mas/pixel and rotation $-0\fdg012$.
The two $\Delta$ columns indicate the distance between that position and the initial {\em HST}-derived position.
This image was taken to verify an already-calibrated WCS, therefore the two sets of $\Delta$ values and the rms $\Delta$ in the last row
are nearly identical.}
\tablenotetext{a}{$R \approx 11$ mag AO guide star, not included in fit due to large $\Delta$.}
\end{deluxetable}

\clearpage

\begin{deluxetable}{rrrrr}
\tablecolumns{5}
\tablewidth{0pc}
\tabletypesize{\scriptsize}
\tablecaption{HIP 62403 Field Astrometric Data \label{tab-HIP62403}}
\tablehead{\colhead{} & \multicolumn{2}{c}{Coordinates} & \multicolumn{2}{c}{Red Channel} \\
\colhead{No.} & \colhead{$\alpha$} & \colhead{$\delta$} & \colhead{$X$} & \colhead{$Y$} \\
}
\startdata
62403 & 191.8289838 & -66.2374484  &   507.5  &  487.5 \\
    1 & 191.8346683 & -66.2355670  &    49.0  &  864.0 \\
    2 & 191.8333292 & -66.2348526  &   157.0  & 1007.0 \\
    3 & 191.8305338 & -66.2394045  &   382.5  &   96.0 \\
    4 & 191.8283639 & -66.2360693  &   557.5  &  763.5 \\
    5 & 191.8277686 & -66.2394120  &   605.5  &   94.5 \\
    6 & 191.8250159 & -66.2385176  &   827.5  &  273.5 \\
    7 & 191.8247250 & -66.2357544  &   851.0  &  826.5 \\
    8 & 191.8245947 & -66.2363066  &   861.5  &  716.0 \\
\enddata
\tablecomments{Derived coordinates for the HIP 62403 field.  The celestial coordinates are in degrees 
for equinox J2000, epoch 2011.37.  HIP 62403 computed for 2011.37 from J2000.0 coordinates plus 
the parallax and proper motion; other stars from their position relative to 62403 using the LMC-11mag WCS.  
The estimated mean error is $6\times10^{-6}$ degrees, dominated by the distortion correction.
The Red channel ($X$,$Y$) represent the image coordinates in pixels after correcting distortion.}
\end{deluxetable}

\clearpage

\begin{deluxetable}{lccccclclcccccclccc}
\scriptsize
\setlength{\tabcolsep}{3pt}  
\tablecolumns{19}
\tablewidth{0pc}
\tabletypesize{\scriptsize}
\rotate
\tablecaption{ NICI Astrometric Field Comparison \label{tab-astromcorr}}
\tablehead{
  \colhead{} & \colhead{} & \multicolumn{6}{c}{Early Dataset} & & \multicolumn{7}{c}{Reference Dataset} & \colhead{} & \colhead{} & \colhead{} \\
  \cline{3-7} \cline{9-16} \\
  \colhead{Target} & \colhead{\# Bgd} & \colhead{Date UTC} & \colhead{Chan} & \colhead{DW} & \colhead{Filter} & \colhead{$\theta_{\rm IAA}$} & &
     \colhead{Date UTC} & \colhead{Chan} & \colhead{DW} & \colhead{Filter} & \colhead{Mounting} & \colhead{Port} & \colhead{$\theta_{\rm IAA}$} & \colhead{Calib.} & 
     \colhead{$\Delta\theta$} & \colhead{rms Error} & \colhead{Note}\\
   &\colhead{Stars}& & & & &\colhead{($^\circ$)} & & & & & & & & \colhead{($^\circ$)}& \colhead{Field} &\colhead{($^\circ$)} & \colhead{(pixels)}
}
\startdata
\multicolumn{9}{l}{Mounting 1, 2008 Jul 27, Port 5 (Side-looking), Calibrator = 70 Oph} \\
HD 61005    &   2   & 2009-01-13 & B    & 50/50  & 4\%S   & 112.30 & & 2011-04-25 & B & Mir   & H    & 6 & 1 & 247.50 & LMC    & 0.215 & 0.23 \\
RXJ 1224.8  &   2   & 2009-02-06 & R    & 50/50  & 4\%L   & 112.30 & & 2012-04-06 & R & H/K   & Ks   & 9 & 5 & 112.64 & LMC    & 0.040 & 0.92 \\
Gamma Oph   &   7   & 2009-04-08 & B    & Mir    & H      & 112.30 & & 2010-04-08 & B & Mir   & H    & 3 & 5 & 112.20 &        & 0.150 & 1.54 & \tablenotemark{a} \\
Prox Cen    &   8   & 2009-04-26 & R    & 50/50  & 4\%L   & 112.30 & & 2009-04-08 & R & 50/50 & 4\%L & 1 & 5 & 112.30 & 70 Oph & 0.478 & 0.39 & \tablenotemark{b} \\
HD 196544   &   2   & 2009-04-26 & R    & 50/50  & 4\%L   & 112.30 & & 2011-04-25 & B & Mir   & H    & 6 & 1 & 247.50 & LMC    & 0.700 &      & \tablenotemark{c} \\
HD 196544   &   2   & 2009-04-27 & R    & 50/50  & 4\%L   & 112.30 & & 2011-04-25 & B & Mir   & H    & 6 & 1 & 247.50 & LMC    & 0.400 \\
\\
\multicolumn{9}{l}{Mounting 2, 2009 Oct 24, Port 1 (Up-looking), Calibrator = HDO 171 B-C} \\
HIP 41373   &\nodata& 2010-01-10 & B    & 50/50  & 4\%S   & 247.50 & & 2011-02-05 & B & Mir   & H    & 5 & 1 & 247.50 & LMC    & 0.274 & 0.93 \\
HIP 62403   &   4   & 2010-02-28 & R    & 50/50  & 4\%L   & 247.50 & & 2011-04-15 & R & 50/50 & 4\%L & 6 & 1 & 247.50 & LMC    & 0.077 & 2.04 & \tablenotemark{d} \\
\\
\multicolumn{9}{l}{Mounting 3, 2010 Mar 20, Port 5 (Side-looking), Calibrator = 70 Oph} \\
70 Oph      &\nodata& 2010-04-07 & R\&B & 50/50  & 4\%L,S \\
HD 168443   &   6   & 2010-04-08 & B    & 50/50  & 4\%L   & 112.15 & & 2012-04-06 & R & 50/50 & 4\%L & 9 & 5 & 112.64 & LMC    & 0.046 &      & \tablenotemark{e} \\
70 Oph      &\nodata& 2010-04-09 & R\&B & 50/50  & 4\%L,S & 112.15 & &            &   &       &      &   &   &        &        &       &      & \tablenotemark{f} \\
70 Oph      &\nodata& 2010-05-09 & R\&B & 50/50  & 4\%L,S & 111.70 & &            &   &       &      &   &   &        &        &       &      & \tablenotemark{g} \\
HIP 59315   &   2   & 2010-05-09 & R    & 50/50  & 4\%L   & 111.70 & & 2011-05-12 & R & H/K   & Ks   & 7 & 1 & 247.50 & LMC    &-0.604 &      & \tablenotemark{h} \\
            &       &            &      &        &        & 112.20 & &            &   &       &      &   &   &        &        & 0.02  & 0.92 & \tablenotemark{i} \\ 
\enddata
\tablecomments{The table compares {\em Early} binary-star based astrometric calibrations to later more 
  reliable {\em Reference} calibrations for several fields observed over multiple epochs.
  For each field, the parameters of both the Early and Reference datasets are listed, followed by $\Delta\theta$ = the
  residual rotation angle between the two images after derotating according to the nominal WCS, and the rms difference
  between the star positions in the two datasets.  Abbreviations are: R: Red channel; B: Blue channel; 
  50/50: H-50/50 beamsplitter; Mir: Mirror; 4\%L, 4\%S: CH4-H4\%L, S filters. }
\tablenotetext{a}{2010 WCS corrected from {\iaa} = 112.15 to 112.20.}
\tablenotetext{b}{2009-04-26 anomalous by 0\fdg478; see \S\ref{sec-earlycheck}}
\tablenotetext{c}{Derived from HD star and 1 background star.}
\tablenotetext{d}{{LMC} field also taken 2011-04-15, {\iaa} confirmed to -0\fdg02.}
\tablenotetext{e}{Dataset {\iaa} = 112.15 is nearly correct.}
\tablenotetext{f}{Derived {\iaa} = 111.7 without distortion correction, 112.4 with correction.}
\tablenotetext{g}{Dataset confirms 112.4 with distortion correction.}
\tablenotetext{h}{{\iaa} = 111.70 is erroneous.}
\tablenotetext{i}{Changing {\iaa} to 112.2 corrects error.}
\end{deluxetable}

\clearpage

\begin{deluxetable}{rccc}
\tablecolumns{4}
\tablewidth{0pc}
\tabletypesize{\footnotesize}
\tablecaption{WCS Correction Parameters \label{tab-wcscorr}}
\tablehead{
  \multicolumn{2}{c}{Mounting} \\
  \colhead{\#} & \colhead{Date UTC} & \colhead{Port} & \colhead{$I\!A\!A_{\rm true}$} \\
               &                    &                &   ($^\circ$)  }
\startdata
 1 & 2008-07-27 & 5 & 112.30 \\
 2 & 2009-10-24 & 1 & 247.50 \\
 3 & 2010-03-20 & 5 & 112.20 \\
 4 & 2010-10-20 & 5 & 112.61 \\
 5 & 2011-01-14 & 1 & 247.50 \\
 6 & 2011-03-11 & 1 & 247.50 \\
 7 & 2011-04-27 & 1 & 247.50 \\
 8 & 2011-06-08 & 5 & 112.82 \\
 9 & 2012-02-15 & 5 & 112.64 \\
10 & 2012-07-13 & 5 & 112.42 \\
11 & 2013-05-07 & 5 & 112.25 \\
12 & 2013-06-19 & 1 & 247.50
\enddata
\tablecomments{{\iaa} values used by the {\tt NICI\_FIXWCS} routine. Pixel Scales used are: Red = 17.958, Blue = 17.994 mas/pixel.}
\end{deluxetable}

\clearpage

\clearpage 

\clearpage

\clearpage

\end{document}